\documentclass[aps,pra,onecolumn,notitlepage]{revtex4}
\usepackage{amsmath, amssymb,physics}
\usepackage{bm}
\usepackage{natbib}
\usepackage{color}

\usepackage{graphicx}
\usepackage{float}

\begin{document}

\title{Revisiting the Acousto-Electric Effect}

\author{E. M. Wright, J. Mack, A. Wendt, A. Burrington, W. Roberts, S. Yusofsani, D. Anderson, and M. Eichefield}
\affiliation{James C. Wyant College of Optical Sciences, University of Arizona, \\ Tucson, Arizona 85721, USA}

\begin{abstract}  
The goal of this paper is to provide a new perspective on the acousto-electric effect by deriving a wave equation for the acoustic field that is akin to Stokes 1845 viscous wave equation and in which the phonon-electron interaction provides the loss/gain term.  We hope this new perspective may provide some insight into the workings of the acousto-electric effect, and we use it to build connections to other areas of research, in particular inertial motion superradiance and the Zel'dovich effect.
\end{abstract}

\maketitle

\section{Introduction and preview}

The acousto-electric (AE) effect, in which an electric current can be generated by a traveling acoustic wave in a piezoelectric semiconductor, was predicted theoretically by R. Parmenter in 1953 \cite{Parm53,Wein56}. The first experimental demonstration of ultrasonic amplification in a piezoelectric semiconductor following in 1961 \cite{HusMcFWhi61}.  The amplification or gain of the acoustic wave, or phonon field, arises from the phonon-electron interaction afforded by the AE effect in conjunction with a bias electric field, $E_0$, that provides a drift velocity $v_d>v_a$ for the electrons, $v_a$ being the velocity of the acoustic mode.  Experimental research on AE amplification continues to be an active area, see Ref. \cite{HacMilWea23}, and it has now expanded to include nonlinear phononics, the realization of surface acoustic wave phonon lasers \cite{Wendt25}, and even quantum acoustoelectric interactions.

It is fair to say that the AE effect is extremely well understood theoretically.  A seminal theory paper was already published in 1962 by White \cite{White62}, and a simplified approach based on circuit theory by Adler appeared in 1971 \cite{Adl71}.  In addition, Kino and Reeder \cite{KinRee71} developed the theory of Rayleigh or surface acoustic wave amplification, and Conwell and Ganguly extended the theory to allow for nonlinear three-wave mixing of acoustic waves \cite{ConGan71}.  The basic theoretical procedure for amplification is to seek solutions for a traveling acoustic wave coupled to the electron system and find conditions for which amplification can appear. This is assuredly the correct physical approach, and a key requirement for acoustic gain is that the drift velocity $v_d$ of the electrons must exceed the acoustic phase velocity $v_a$. Adler points out that if the acoustic frequency in the lab frame is $\omega=v_ak_a$ , then the acoustic wave frequency seen by electrons moving at velocity $v_d$ is Doppler shifted to $\omega_a =\omega(1-v_d/v_a)=\omega-k_a v_d$, which becomes negative if $v_d > v_a$ \cite{Adl71}.  Acousto-electric amplification is therefore intimately related to the appearance of negative frequencies, albeit in the reference frame moving at the electron drift velocity. The detailed gain also depends on the time scale $\tau$ at which the electron system relaxes to equilibrium, and transparency or zero gain occurs for $v_d=v_a$.

Given that the AE effect is so well understood, this begs the question of the relevance of the present paper.  For sure the presentation here will not change any analysis or calculation of AE effect based devices, but what we do here is revisit the AE effect and provide a different perspective that reveals ties to other research areas.  To proceed, we now give a preview of our main result as a guide to the reader: in our approach we develop a wave equation for the material displacement $u(x,t)$ describing a one-dimensional acoustic field without explicit reference to the density of electrons, but which accounts for the effects of the phonon-electron interactions.  As a reference point, in the field of acoustics (i.e. a medium that can only support longitudinal strain such as an ideal gas or liquid), a wave equation incorporating losses takes the generic form
\begin{equation}
{\partial^2 u\over \partial t^2} - v_a^2{\partial^2 u\over \partial x^2} = \Gamma {\partial^2\over \partial x^2} {\partial u \over \partial t }, 
\end{equation}
where $\Gamma$ is the viscous damping coefficient.  This is the famous Stokes viscous acoustic wave equation published in 1845 and derived from the Navier-Stokes equations \cite{Stokes}, and it finds application in diverse fields such as seismic waves, ultrasound imaging, and microfluidics. For the case of AE interactions, we will show the right-hand-side is due to the phonon-electron coupling and describes loss and gain.  We show in this paper that this wave equation can be applied to the AE effect in the vicinity of transparency with two caveats.  First, since in our case the losses are due to the electrons that can be drifting at velocity $v_d$ with respect to the frame of the acoustic waves, the time derivative on the right-hand-side needs to be replaced by the convective derivative ${D\over Dt} = {\partial\over\partial t} + v_{d}{\partial\over \partial x}$.  Second, the viscous damping coefficient is given by $\Gamma=v_a^2K^2\tau$, where $K^2<<1$ is the square of the electromechanical coupling coefficient.  Bringing this together yields
\begin{equation}\label{Stokes}
{\partial^2 u\over \partial t^2} - v_a^2{\partial^2 u\over \partial x^2} = v_a^2K^2\tau {\partial^2\over \partial x^2} \left  ( {\partial \over\partial t} + v_{d}{\partial \over \partial x} \right )u .
\end{equation}
To get a sense of how this leads to AE amplification, it is useful to consider a monochromatic traveling wave solution of the form $u(x,t)= u_0e^{i(k_a x -\omega t)}$, the convective derivative of which produces the prefactor $(\omega-k_av_d)$ multiplying the loss.  We therefore see that the convective derivative in Eq. (\ref{Stokes}) naturally produces the factor $\omega(1-v_d/v_a)$ that describes the crossover from loss $(v_d<v_a)$ to gain $(v_d>v_a)$ alluded to above.  It is our opinion that approaching the AE effect starting from such a common equation from acoustics might help make the transition from acoustic loss to gain seem less mysterious: Physically, loss of acoustic energy must be balanced by increase in the electron kinetic energy (heating), and gain of acoustic energy must be balanced by decrease in the electron kinetic energy (cooling).

The remainder of this paper is aimed at deriving Eq. (\ref{Stokes}) using the one-dimensional model for AE amplification in Ref. \cite{White62}. To the best of our knowledge, the approach is new, and although not technically difficult we hope it offers some new insight.  Following this we discuss how the perspective offered here can bridge connections to other research areas, in particular inertial motion superradiance \cite{BecSch98,Brito} and the Zel'dovich effect \cite{Zeldy} for acoustic waves.

\section{Governing equations}

Our goal in this Section is to first obtain the coupled equations that govern AE interactions in a piezoelectric semiconductor using the model of Ref. \cite{White62}.  Using the coupled equations we then derive the effective acoustic wave equation and verify that it yields the same expression for the AE amplification as Adler's simplified model \cite{Adl71}.

\subsection{Constitutive relations}

We first examine the constitutive relations of piezoelectric materials, which describe the coupling between electrical and mechanical properties
\begin{eqnarray}\label{ConsEqs}
T &=& c S - e E,  \nonumber \\
D &=& e S + \varepsilon E ,
\end{eqnarray}
where we have followed the notation in Ref. \cite{White62}, and $T$ is the elastic stress field, $S$ the elastic strain field, $c$ the material elasticity (evaluated under a constant electric field), $E$ the electric field, $e$ the piezoelectric constant, $D$ is the electric displacement field, and $\varepsilon$ is the dielectric permittivity (evaluated under a constant strain field). We can rewrite the expression for $D$ in Eq. (\ref{ConsEqs}) in terms of the material displacement field $u(x,t)$ using the relation $S = {\partial u\over \partial x}$ to obtain
\begin{equation}\label{D}
D = e \pdv{u}{x} + \varepsilon E ,
\end{equation}
and taking the partial spatial derivative of Eq. (\ref{D}) yields
\begin{equation}\label{PD}
\pdv{D}{x} = e \pdv[2]{u}{x} + \varepsilon \pdv{E}{x} .
\end{equation}
Next we employ Gauss' law in one dimension, $\pdv{D}{x} = -q(n_c-n_0)$, where $n_c$ is the electron density. More specifically, we set 
\begin{equation}\label{nc}
n_c(x,t) = n_0 + n_s(x,t) ,
\end{equation}
where $n_0$ accounts for the equilibrium density of electrons and $n_s(x,t)$ is a contribution due to coupling to the acoustic field. Substituting Gauss's law into Eq. (\ref{PD}) we then obtain
\begin{equation}\label{rho1}
Q = -qn_s = e \pdv[2]{u}{x} + \varepsilon \pdv{E}{x}, 
\end{equation}
$Q$ being the space charge.

\subsection{Acoustic wave equation}

To proceed, we use the piezoelectric wave equation from Ref. \cite{White62}
\begin{equation}\label{White}
\rho \pdv[2]{u}{t} = {\partial T\over \partial x} =c \pdv[2]{u}{x} - e \pdv{E}{x},
\end{equation}
with $\rho$ the elastic medium density.  Then assuming the electric field can be decomposed into a constant component $E_0$ (due to an external bias) and an oscillating component $E_1$ (from the piezoelectric wave)
\begin{equation}\label{E}
E(x,t) = E_0 + E_1(x,t),
\end{equation}
Eq. (\ref{White}) becomes
\begin{equation}\label{Dens}
\rho \pdv[2]{u}{t} = c \pdv[2]{u}{x} - e \pdv{E_1}{x} ,
\end{equation}
and Eq. (\ref{rho1}) may be rearranged as
\begin{equation}\label{Dens2}
\pdv{E_1}{x} = - {1\over\varepsilon} \left (qn_s + e \pdv[2]{u}{x} \right ).
\end{equation}
Combining Eqs. (\ref{Dens}) and (\ref{Dens2}), we obtain the AE wave equation in the form that we shall use below
\begin{equation}\label{WavEq}
\rho \pdv[2]{u}{t} = c \pdv[2]{u}{x} + \frac{e}{\varepsilon} \left( qn_s + e \pdv[2]{u}{x} \right) .
\end{equation}
Note that the term proportional to the electron density $n_s$ on the right-hand-side of this equation acts as a source for the acoustic field.

\subsection{Carrier density equation}

If the electron collision frequency $1/\tau_c$, $\tau_c$ being the carrier collision time, is much larger than the frequency $\omega$ of the acoustic wave, the current density $J$ in an $n$-type semiconductor is given by the convection-diffusion equation \cite{White62}
\begin{equation}\label{J}
J = \mu q n_c E + q D_n \pdv{n_c}{x} ,
\end{equation}
where $q$ is the magnitude of the electron charge, $\mu$ is the electron mobility, $n_c = n_0 + n_s$ as in Eq. (\ref{nc}), and $D_n$ is the electron-diffusion coefficient.  We note that this expression for the current density contains the product
\begin{equation}\label{ncE}
n_c(x,t) E(x,t) = n_0E_0 + \underbrace{ E_0 n_s(x,t) + n_0 E_1(x,t) } + E_1(x,t)n_s(x,t)  ,
\end{equation}
where Eqs. (\ref{nc}) and (\ref{E}) have been used.  In the small signal approximation that we assume here $|E_0| >> |E_1|$ and $n_0>> |n_s|$, meaning that the last term may be dropped to leading order and the underbraced terms are the relevant terms, the first term being a constant. The last term gives rise to a nonlinear current
\begin{equation}\label{JNL}
J^{NL}=\mu q n_s E_1,
\end{equation}
that will produce harmonics of the acoustic field and we drop for the time being.  Then using the charge continuity equation $\pdv{Q}{t} + \pdv{J}{x} = 0$ along with $Q = -q n_s$, and in the small signal approximation, we obtain
\begin{equation}\label{dJdx}
    \pdv{J}{x} =  -q \pdv{n_s}{t} = \mu q E_0\pdv{n_s}{x} + \mu 
    q n_0\pdv{E_1}{x} + q D_n \pdv[2]{n_s}{x} ,
\end{equation}
which upon rearrangement and use of Eq. (\ref{Dens2}) yields
\begin{equation}
\pdv{n_s}{t} = -\frac{n_0 \mu}{\varepsilon} \left( qn_s + e\pdv[2]{u}{x} \right) + \mu E_0 \pdv{n_s}{x} + D_n \pdv[2]{n_s}{x}.
\end{equation}
Using the definition $\mu = q \tau_c / m^*$ along with the plasma frequency $\omega_p^2 = n_0 q^2 / \varepsilon m^*$, $m^*$ being the effective mass of the electron, we obtain
\begin{equation}\label{DensEq}
    \pdv{n_s}{t} - \mu E_0 \pdv{n_s}{x} = -\omega_p^2 \tau_c \left( n_s +\frac{e}{q} \pdv[2]{u}{x} \right) + D_n \pdv[2]{n_s}{x}.
\end{equation}
Note that the term proportional to the second derivative of the material displacement $u(x,t)$ on the right-hand-side of this equation acts as a source for the electron density.  In addition, the left-hand-side has the form the convective derivative ${D\over Dt} = {\partial\over\partial t} + v_{d}{\partial\over \partial x}$ where $v_d=-\mu E_0$, the first term on the right-hand-side proportional to $n_s$ describes electron relaxation with rate ${1\over\tau}=\omega_p^2 \tau_c$, $\tau$ being the time scale on which the electron system returns to equilibrium, and the last term describes electron diffusion.  In what follows we shall neglect the last term in the basis that the electron diffusion occurs on a time scale longer than the $\tau$.

\subsection{Coupled equations}
We now give the coupled equations for the acoustic field and density of electrons in a form suitable for analysis.  Starting from the acoustic wave equation in Eq. (\ref{WavEq})
\begin{equation}\label{Eq1}
\rho {\partial^2 u\over \partial t^2} = \left ( c+ {e^2\over \varepsilon} \right ){\partial^2 u\over \partial x^2} + {eq\over \varepsilon} n_s,
\end{equation}
it is useful to rearrange this as
\begin{equation}\label{Eq3}
{\partial^2 u\over \partial t^2}  - v_a^2 {\partial^2 u\over \partial x^2} = v_a^2 K^2 {\partial^2 u\over \partial x^2} + {eq\over \rho\varepsilon} n_s,
\end{equation}
where $v_a=\sqrt{{c\over\rho}}$ is the acoustic velocity with no coupling to the carrier density, and $K^2={e^2\over \varepsilon c}<<1$ is the square of the electromechanical coupling coefficient.  The left-hand-side of Eq. (\ref{Eq3}) is the acoustic wave equation with no coupling to the charge carriers, and the right-hand-side describes the coupling to the carrier density which is weak under the assumption $K^2<<1$.  In the absence of diffusion we next write the electron density Eq. (\ref{DensEq}) as 
\begin{equation}\label{Eq4}
{n_s\over\tau}  + {\partial n_s \over \partial t } +v_d {\partial n_s \over \partial x} = - {\mu en_0\over \varepsilon} {\partial^2 u\over \partial x^2},
\end{equation}
where as before ${1\over\tau} = (\omega_p^2\tau_c)$ is the rate at which the electron density relaxes back to equilibrium, and $v_d=-\mu E_0$, the negative sign being due to the negative charge of the electron. 

\section{Effective acoustic wave equation}

\subsection{Acoustic attenuation}

The idea is to obtain an approximate solution for $n_s$ from Eq. (\ref{Eq4}) that can be used in the acoustic wave Eq. (\ref{Eq3}) to obtain an effective acoustic wave equation. The key assumption is that the dominant term on the left-hand-side of Eq. (\ref{Eq4}) is the first one proportional to ${1\over\tau}$, and we rearrange as
\begin{equation}\label{Eq5}
{1\over\tau} \left [ 1 +\tau\left ( {\partial \over \partial t } + v_d  {\partial \over \partial x} \right ) \right ]n_s= - {\mu en_0\over \varepsilon} {\partial^2 u\over \partial x^2}.
\end{equation}
This can be inverted to yield the formal solution for the electron density
\begin{eqnarray}\label{Eq6}
n_s &=&  -  {\mu en_0\tau\over \varepsilon}\left [ 1 +\tau\left ( {\partial \over \partial t } + v_d {\partial \over \partial x} \right ) \right ]^{-1}{\partial^2 u\over \partial x^2}  \nonumber \\
&=&  -  {\mu en_0\tau\over \varepsilon}{\partial^2 u\over \partial x^2} + {\mu en_0\tau^2\over \varepsilon}\left ( {\partial \over \partial t } + v_d {\partial \over \partial x} \right ){\partial^2 u\over \partial x^2} + \ldots
\end{eqnarray}
where the first two terms of a power series expansion of the solution are shown in the second line.  Substituting the full power series expansion into Eq. (\ref{Eq3}), and using the definitions of the various quantities, we obtain the {\it effective acoustic wave equation}
\begin{eqnarray}\label{Eq7}
{\partial^2 u\over \partial t^2}  - v_a^2 {\partial^2 u\over \partial x^2} &=&  v_a^2 K^2 \left ( 1 - \left [ 1 +\tau\left ( {\partial \over \partial t } + v_d {\partial \over \partial x} \right ) \right ]^{-1}\right ) {\partial^2 u\over \partial x^2} \nonumber \\
&=&  v_a^2 K^2 \sum_{n=1}^\infty (-1)^{n+1} \tau^n \left ( {\partial \over \partial t } + v_d {\partial \over \partial x} \right )^n  {\partial^2 u\over \partial x^2}.
\end{eqnarray}
We remark that it is not uncommon in the treatment of attenuation and dispersion in acoustics that the loss is described by higher-order spatial and temporal derivatives acting on the mechanical displacement field.

\subsection{Acoustic dispersion relation}

As a validation of the effective acoustic wave equation above we here verify that it leads to the same acoustic loss expression as previous work.  In particular, we seek a monochromatic solution for the mechanical displacement of the form $u(x,t)=u_0e^{i(kx-\omega t)}$, and thereby obtain the dispersion relation that yields the (complex) wave vector $k=k'+ik''$ in terms of the acoustic frequency $\omega$ and the parameters of the problem.  Substituting this solution into the first line of Eq. (\ref{Eq7}) yields
\begin{equation}\label{Eq8}
-\omega^2 + k^2 v_a^2  = -k^2 v_a^2 K^2 \left ( 1 - \left [ 1 +\tau\left ( -i\omega + i k v_d \right ) \right ]^{-1}\right ).
\end{equation}
If we set $K=0$, so there is no coupling between the acoustic wave and the carrier density, then the right-hand-side of Eq. (\ref{Eq8}) is zero and we obtain the dispersion relation $k'=k_a =\omega/v_a$, which is real.  Since it is true in experiments that $K^2<<1$, it must be true that even with coupling that the dominant contribution to $k$ must be $k_a$.  This means that we can apply perturbation theory to Eq. (\ref{Eq8}) and on the right-hand-side we can approximate $k\approx k_a=\omega/v_a$.  We then obtain
\begin{equation}\label{Eq9}
-\omega^2 + k^2 v_a^2  = -k_a^2 v_a^2 K^2 \left ( 1 - \left [ 1 -i\omega \tau ( 1 - v_d/v_a) \right ]^{-1} \right ),
\end{equation}
or upon more rearranging
\begin{equation}
k^2 = k_a^2 \left [ 1 + K^2 { i\omega\tau\gamma\over (1-i\omega\tau\gamma)} \right ],
\end{equation}
where $\gamma = (1-v_d/v_a)$.  Using the fact that $K^2<<1$ we Taylor expand the above result to obtain
\begin{equation}
k \approx k_a \left [ 1 + {K^2\over 2}  { i\omega\tau\gamma\over (1-i\omega\tau\gamma)} \right ].
\end{equation}
We can now evaluate the real acoustic wave vector
\begin{eqnarray}\label{kprime}
k'(\omega) &\approx&  k_a - {k_aK^2\over 2}  {(\omega\tau\gamma)^2\over (1+(\omega\tau\gamma)^2)} \nonumber \\
&=& k_a  + \delta k(\omega) ,
\end{eqnarray}
where $k_a={\omega\over v_a}$ and the shift in the real part of the wave vector is given by
\begin{equation}\label{Delta_k}
\delta k(\omega) = - {k_a K^2\over 2} { (k_a\tau)^2(v_a-v_d)^2\over (1+(k_a\tau)^2 (v_a-v_d)^2)} .
\end{equation}
In general $|\delta k|<< k_a$, but this shift will be important when considering phase-matching of nonlinear processes.  In addition, using the fact that the acoustic absorption $\alpha=k''$ we obtain
\begin{eqnarray}\label{alpha}
\alpha &\approx&  {k_aK^2\over 2}  {\omega\tau\gamma\over (1+(\omega\tau\gamma)^2)} \nonumber \\
&=&  {k_a K^2\over 2}  {(k_a\tau)(v_a-v_d)\over (1+(k_a\tau)^2 (v_a-v_d)^2)} .
\end{eqnarray}
This expression for the absorption coincides with that given in the papers by White \cite{White62} and Adler \cite{Adl71}.  Imposing the further condition $|k_a\tau (v_a-v_d)| << 1$ implies operating close to transparency as the expression for the loss reduces to
\begin{equation}\label{Eq10}
\alpha \approx {k_a^2K^2\tau\over 2} (v_a-v_d).
\end{equation}
We shall consider this limit around transparency, which corresponds to taking the first term in the expansion in Eq. (\ref{Eq7}), as it allows us to focus on the transition between loss $(v_a > v_d)$ and gain $(v_d > v_a)$.  In general, retention of the neglected higher-order terms allows for the treatment of dispersive effects both in terms of the refractive index and gain or loss of the acoustic waves.
\section{Physical interpretation}

To gain some insight into the underlying physics involved in the transition from loss to gain we retain only the first term $(n=1)$ from the power series expansion in Eq. (\ref{Eq7}) to obtain
\begin{equation}\label{Eq11}
{\partial^2 u\over \partial t^2} - v_a^2 {\partial^2 u\over \partial x^2}  = v_a^2 K^2 \tau\left ( {\partial \over \partial t } + v_d {\partial \over \partial x} \right ) {\partial^2 u\over \partial x^2}.  
\end{equation}
Based on the discussion of the previous Section this resulting equation is applicable in the vicinity of transparency, and is Eq. (\ref{Stokes})  alluded to in the Introduction.  In the case with no drift, $v_d=0$, this has the form of Stokes acoustic wave equation with attenuation
\begin{equation}
{\partial^2 u\over \partial t^2} - v_a^2{\partial^2 u\over \partial x^2} = \Gamma {\partial^2\over \partial x^2} {\partial u \over \partial t }, 
\end{equation}
with $\Gamma=v_a^2K^2\tau$ the acoustic wave attenuation coefficient.  This form of the wave equation for sound with attenuation is called the viscous wave equation and was first derived by Stokes in 1845 \cite{Stokes}. The extra spatial derivative term that appears on the right-hand-side of Eq. (\ref{Eq11}) is what is expected if the acoustic wave is propagating in a moving and absorbing acoustic medium, in this case the background electron drift motion due to the bias electric field.

We contend that acousto-electric amplification is an example of the phenomenon of {\it superradiance} which refers to the amplification of waves in a medium via scattering from a moving object \cite{BecSch98,Brito}.  Inertial motion superradiance arises for objects moving at constant velocity with respect to the  medium that is stationary in the lab reference frame, whereas rotational superradiance occurs arises for rotating objects as the name suggests.  Here we focus on the case of inertial motion superradiance.  For our particular case the waves are acoustic waves of frequency $\omega=k_av_a$ in the lab frame and the moving object is the electron current stream which is moving at the drift velocity $v_d$ by virtue of the bias electric field.  In the rest frame of the moving electrons the acoustic frequency is thus Doppler shifted to $\omega'=\omega-k_a v_d$, and the criterion for superradiance is that $\omega'< 0$: That is, the drift velocity is large enough that the acoustic frequency becomes negative in the electron rest frame, this possibility being referred to as the anomalous Doppler effect.  From a quantum perspective the phonon energy in the reference frame moving at the drift velocity is $\hbar\omega'$, that is the waves have negative energy!  Then if the drifting electrons absorb this negative energy phonon they must reduce their mean kinetic energy and hence temperature.  Back in the lab frame the acoustic field can then appear amplified, the added energy being balanced by the reduction in the temperature and drift velocity of the electrons.  This picture is the basis of our treatment of gain saturation given in the next Section.

We mention that rotational superradiance has previously been explored both theoretically \cite{Zeldy,BecSch98,FacWri19} and experimentally \cite{News20,CroGib20} in acoustics.  In this case, an acoustic vortex beam is incident at right angles to a rotating and absorbing medium and amplification is seen on transmission through the medium, along with concomitant negative frequencies.  The analysis given in those papers makes use of the same Stokes viscous wave equation discussed here.  Rotational superradiance is also referred to as the Zel'dovich effect \cite{Zeldy,FacWri19}, and it has analogies to scattering of waves from rotating black holes \cite{News20,CroGib20}. It would be interesting to see if rotational superradiance could also be observed using the AE effect with acoustic vortices and current rings.

\section{Gain saturation mechanism}

In this Section we discuss a gain saturation mechanism that is intimately tied to the above interpretation of the AE amplification as due to superradiance \cite{BecSch98}.  It does require the addition of a thermo-acoustic equation so that the balance between the energy of the acoustic field and the electron temperature can be incorporated.

\subsection{Acoustic wave propagation}

Here we start from the wave Eq. (\ref{Stokes})
\begin{equation}
{\partial^2 u\over \partial t^2} - v_a^2{\partial^2 u\over \partial x^2} = v_a^2K^2\tau {\partial^2\over \partial x^2} \left  ( {\partial \over\partial t} + v_{d}{\partial \over \partial x} \right )u ,
\end{equation}
and we substitute the following ansatz for an acoustic wave traveling along the positive x-axis
\begin{equation}
u(x,t) = U(x)e^{i(k_a x -\omega t)},
\end{equation}
where $U$ is the slowly varying envelope of the acoustic field.  In the slowly varying envelope approximation we then obtain
\begin{equation}
-2ik_a v_a^2 {dU\over dx} = -v_a^2K^2\tau k_a^2 (-i\omega + i k_a v_d) U,
\end{equation}
or rearranging
\begin{equation}
{dU\over dx} = -\alpha U
\end{equation}
where the acoustic loss is
\begin{equation}\label{loss2}
\alpha = {k_a^2K^2\tau\over 2} (v_a-v_d).
\end{equation}
which agrees with Eq. (\ref{Eq10}) in the vicinity of transparency.  (For simplicity in presentation we assume that both $v_a$ and $v_d$ are positive).  We also introduce the acoustic energy flux or intensity
\begin{equation}
I = {1\over 2} \rho\omega^2 v_a |U|^2,
\end{equation}
which obeys Beer's law under propagation
\begin{equation}
I(x) = I(0)e^{-2\alpha x}  ,
\end{equation}
under the assumption that the intensities are small enough to neglect saturation.

\subsection{The drift velocity}

We start from Newton's equation for the electron motion

\begin{equation}
m^* {dv\over dt} = -qE_0 - {m^* v\over \tau_c} ,
\end{equation}
where $v={dx\over dt}$, and $\tau_c$ is the carrier collision time.  From this we evaluate the time rate of change of the electron kinetic energy

\begin{equation}
{dE_{kin}\over dt} = {d\over dt} \left [ {m^*\over 2} \left ( {dx\over dt} \right )^2 \right ] = -qE_0v - {m^* v^2\over \tau_c} ,
\end{equation}
and in steady-state we then obtain 
\begin{equation}
v = v_d^{(0)} = -{qE_0\tau_c\over m^*} = -\mu E_0,
\end{equation}
with $\mu=q\tau_c/m^*$ the electron mobility as before.  Here $v_d^{(0)}$ is the leading order contribution to the drift velocity with no acoustic field present.  We may also assign a temperature to this drift velocity using
\begin{equation}
k_B {\cal T}_d = {m^* (v_d^{(0)})^2\over 2} = {q^2E_0^2 \tau_c^2\over 2 m^*},
\end{equation}
this being the mean kinetic energy per electron with $k_B$ Boltzmann's constant.

\subsection{Thermo-acoustic effect and gain saturation}

We note that the time scale for the drift velocity and associated temperature above to be established is very short, on the order of $\tau_c$ which can be picoseconds or shorter.  As such these quantities are not dependent on effects such as electron diffusion that occur on much longer time scales.  On longer time scales the acoustic wave can modify the electron temperature, and the appropriate equation for the temperature change $\Delta{\cal T}$ for our case can be gleaned from thermo-acoustics as
\begin{equation}
\rho c_v \pdv{\Delta {\cal T}}{t} = \kappa\nabla^2 \Delta {\cal T} + 2\alpha I,
\end{equation}
where $\rho$ is the elastic medium density, $c_v$ the heat capacity per unit mass, and $\kappa$ the thermal conductivity.  The first term on the right-hand-side describes thermal diffusion along with suitable boundary conditions, and the seconds term describes the effect of the acoustic field:  For loss $\alpha >0$ the second term describes heating, whereas for gain $\alpha < 0$ cooling of the electron temperature occurs.  This is the origin of the proposed gain saturation mechanism.

In the spirit of the simple model used in this paper we shall reduce the thermo-acoustic equation to
\begin{equation}
\pdv{\Delta {\cal T}}{t} = -{\Delta {\cal T}\over \tau_{diff}}  + {2\alpha I\over \rho c_v} ,
\end{equation}
where $\tau_{diff}$ is a phenomenological heat decay time that will depend on the material properties as well as the boundary conditions involved.  Then in steady-state we obtain
\begin{equation}
\Delta {\cal T} = {2\tau_{diff}\alpha I\over \rho c_v} .
\end{equation}
We may now extract the correction $v_d^{(1)}$ to the drift velocity using
\begin{equation}
k_B ({\cal T}_d + \Delta{\cal T}) = {m^*\over 2}\left ( v_d^{(0)}+ v_d^{(1)} \right )^2 \approx k_B {\cal T}_d + m^*v_d^{(0)}v_d^{(1)}.
\end{equation}
Using this equation to first-order we then find
\begin{equation}\label{EqXX}
v_d^{(1)}  = {2k_B \tau_{diff}\over m^*v_d^{(0)} \rho c_v} \alpha I.
\end{equation}
We can now see the gain saturation mechanism at work: Consider $v_d > v_a$ so the prefactor $(1-v_d/v_a)$ in the loss expression is negative and there is gain.  As a result of the gain $\alpha<0$ the acoustic intensity $I$ will grow, and the correction to the drift velocity $v_d^{(1)}$ will become more negative, so the net drift velocity $v_d=v_d^{(0)}+v_d^{(1)}$ will become smaller and the gain will decrease according to the prefactor $(1-v_d/v_a)$, that is, the gain will saturate as the acoustic intensity increases.

To be more precise, we can extract the first-order change in the drift velocity using Eq. (\ref{EqXX}) expressed as
\begin{eqnarray}
v_d^{(1)}  &=& {2k_B \tau_{diff}\over m^*v_d^{(0)} \rho c_v} \cdot {k_a^2K^2\tau\over 2} \left ( v_a-v_d \right )I \nonumber \\
&=& (v_a-v_d^{(0)} - v_d^{(1)} ){I\over I_{sat}} ,
\end{eqnarray}
where the saturation intensity is given by
\begin{equation}
{1\over I_{sat}} = {2k_B \tau_{diff}\over m^*v_d^{(0)} \rho c_v} \cdot {k_a^2K^2\tau\over 2} .
\end{equation}
We can now solve explicitly for the first-order change in the drift velocity
\begin{equation}\label{vd1}
v_d^{(1)} = {(v_a-v_d^{(0)})(I/I_{sat})\over 1+I/I_{sat}} 
\end{equation}
Finally, we use Eq. (\ref{loss2}) for the acoustic loss to obtain
\begin{eqnarray}\label{loss3}
\alpha &=&  {k_a^2K^2\tau\over 2} (v_a-v_d^{(0)} - v_d^{(1)}) \nonumber \\
&=&  \underbrace{{k_a^2K^2\tau\over 2} (v_a - v_d^{(0)}) }_{\alpha_0} {1\over (1+I/I_{sat})}.
\end{eqnarray}
Here the underbraced term $\alpha_0$ describes the low intensity or small signal loss or gain, depending on its sign, and the second term describes the saturation.  This is the key result of this Section.  The propagation equation for the acoustic intensity then becomes
\begin{equation}
{dI\over dx} = -2\alpha I = - 2(\alpha_0 I_{sat})  {I/I_{sat}\over (1+I/I_{sat})}.
\end{equation}
This equation has an analytic solution in terms of the Lambert W function and the variation of the acoustic intensity $I(x)$ versus $x$ can be evaluated for given values of $\alpha_0$ and $I_{sat}$, and the input acoustic intensity $I(0)$.

\subsection{Current saturation}

Now that we have the drift velocity approximation we can evaluate the current saturation.  In particular, to first-order the current is
\begin{equation}
J \approx -qn_0 v_d = -qn_0 (v_d^{(0)} + v_d^{(1)}),
\end{equation}
and using Eq. (\ref{vd1}) we find
\begin{equation}
J = -qn_0 {[v_d^{(0)} + v_a (I/I_{sat})]\over 1+I/I_{sat}} .
\end{equation}
This expression shows that $J$ saturates at $J_{sat} = -qn_0 v_a$ for $I >> I_{sat}$.  Rearranging we then obtain
\begin{eqnarray}
{J\over J_{sat}} &=& {[(v_d^{(0)}/v_a)  +  (I/I_{sat})]\over 1+I/I_{sat}} \nonumber \\
&\approx& (v_d^{(0)}/v_a) - [(v_d^{(0)}/v_a) - 1](I/I_{sat}) , \quad (I/I_{sat}) << 1,
\end{eqnarray}
from which we see that for the case of gain $v_d^{(0)} > v_a$ the current decreases with increasing acoustic intensity $I$.  The physical ideas used here to describe saturation are in keeping with the more general approach in the 1966 paper by Ozaki and Mikoshiba \cite{OzaMik66}.

\section{Summary}

We have revisited the acousto-electric effect with the goal of presenting a new physical perspective of the phenomenon.  In particular, for operation near transparency, where the net loss/gain is zero, we were able to recast the acoustic wave equation in the form of Stokes 1845 viscous wave equation.  In particular, as the drift velocity exceeds the acoustic velocity the loss associated with the viscous damping can be seen to turn into gain.  By recognizing the relation between this gain mechanism and inertial frame superradiance we were able to build on this to develop a theory of gain saturation built on the notion that the carriers must be cooled in order to maintain energy conservation as the acoustic wave energy is amplified.  We hope some readers will derive some physical insight from these developments.

\end{document}